Title: Parallel communicating one-way reversible finite automata system

Authors: Debayan Ganguly[1], Kingshuk Chatterjee [2], Kumar Sankar Ray[3]

Affiliations:
[1]Government College of Engineering and Leather technology, Kolkata-700106
[2] Government College of Engineering and Ceramic technology, Kolkata-700010
[3]Electronics and Communication Sciences Unit, Indian Statistical Institute, Kolkata-108.

# Parallel communicating one-way reversible finite automata system

*Debayan ganguly[1], Kingshuk Chatterjee [2], Kumar Sankar Ray[3]*

[1]Government College of Engineering & Leather Technology, Kolkata-700106
[2] Government College of Engineering and Ceramic technology,73, Kolkata-700010
[3]Electronics and Communication Sciences Unit, Indian Statistical Institute, Kolkata-700108.

[1]debayan3737@gmail.com,[2]kingshuk.chatterjee@gcect.ac.in,[3]ksray@isical.ac.in

*Abstract: In this paper, we discuss the computational power of parallel communicating finite automata system with 1-way reversible finite automaton as components. We show that unlike the multi-head one way reversible finite automata model (where we are still not sure whether it accepts all the regular languages) parallel communicating one-way reversible finite automata systems can accept all the regular languages. Moreover for every multi-head one way reversible finite automaton there exist a parallel communicating one-way reversible finite automata system which accepts the same language. We also make an interesting observation that although the components of the system are reversible the system as a whole is not reversible. On the basis of which we conjecture that parallel communicating one-way reversible finite automata systems may accept languages not accepted by multi-head one way reversible finite automata.*

## 1. INTRODUCTION

Ever since Bennet [1] showed that reversible Turing machine has the same computational power as a Turing machine; the interest in finding the computational power of reversible versions of other restricted forms of automata has increased significantly. Reversible automata are information preserving machines, thus reversibility enables us to analyze the behavior of the automaton more accurately. The computational powers of two-way reversible models were found to be equal to their deterministic counter parts in most cases [2]. The problem arises when we consider one way variants of the reversible models [3,4]. One-way reversible finite automata cannot accept all regular languages [5]. One-way multi-head reversible finite automata with two heads can accept all unary regular languages [3] but whether one-way multi-head reversible finite automata model accepts all regular languages or not is unknown to us. In this paper, we show that parallel communicating one-way reversible finite automata systems accept all the regular languages.

In parallel communicating finite automata system [6] several finite automata (component) work together and communicate on request by using special query states. When a component i needs information about the state of component j then component i goes to query state $K_j$. The current state of component j is communicated to component i. There are many variants of this parallel communicating model. If a designated component known as the master can only make the queries about the state of other components then the model is centralized. Moreover, if the component j goes back to its start state after replying to the query then the model is returning. In this paper by parallel communicating finite automata systems we mean the non centralized and non returning model. Whenever we use other variants, we explicitly mention them. Every component of the system has its own tape; same input string in all of them; all the components parse from left to right and are initialized to their start state. The system accepts a string if all the components reach their respective final state at the end of computation. The system accepts a string if all the components reach their respective final state at the end of computation. Mitrana et.al [6] discussed the computational power of parallel communicating finite automata system and also showed that non-deterministic/deterministic multi-head finite automata have the same computational power as deterministic/non-deterministic parallel communicating finite automata systems. They have also worked on the pushdown version of parallel communicating system [7]. A detailed survey of parallel communicating finite automata system can be found in [8].

The main claims of this paper are as follows:
- We show that centralized parallel communicating one-way reversible finite automata systems can accept all regular languages. (See Section 4.)
- We also show that for every multi-head one way reversible finite automaton there exist a parallel communicating one-way reversible finite automata system which accepts the same language. (See Section 4.)
- We also make a interesting observation that unlike the existing parallel communicating models where if the components are deterministic/non-deterministic the system is also deterministic/non-deterministic, in case of the reversible variant although the components are reversible the system as a whole is not reversible. (See Section 4.)

## 2. BASIC TERMINOLOGY

The symbol $V$ denotes a finite alphabet. The set of all finite words over $V$ is denoted by $V^*$, which includes the empty word $\lambda$. The symbol $V^+ = V^* - \{\lambda\}$ denotes the set of all non-empty words over the alphabet $V$. For $w \in V^*$, the length of $w$ is denoted by $|w|$. Let $u \in V^*$ and $v \in V^*$ be two words and if there is some word $x \in V^*$, such that $v = ux$, then $u$ is a prefix of $v$, denoted by $u \leq v$. Two words, $u$ and $v$ are prefix comparable denoted by $u \sim_p v$, if $u$ is a prefix of $v$ or vice versa.

### 2.1 One way reversible Finite automata

A partially defined finite automaton is a 5-tuple of the form $M=(Q,V,q_0,F,\delta)$ where $V$ is an alphabet set, the symbol $Q$ denotes the set of states, $q_0$ is the initial state and $F \subseteq Q$ is the set of final states. The function $\delta$ contains a finite number of transition rules of the form $\delta(q,x)=q'$ which denotes that the machine in state $q$ parses $x$ in its input tape and goes to state $q'$ where $x \in V \cup \{\lambda\}$ and $q,q' \in Q$. A partially defined finite automaton is **deterministic** if there are no two transitions of the form $\delta(q,x)=q'$ and $\delta(q,y)=q''$ where $x$ and $y$ are prefix comparable.

A partially defined deterministic finite automaton is **reversible** if each letter induces a partial one-to-one map from the set of states into itself. i.e. if the automaton has a transition of the form $\delta(q,x)=q'$ then it cannot have any other transition of the form $\delta(q'',y)= q'$ where $x,y \in V \cup \{\lambda\}$ and $q,q',q'' \in Q$ where $x$ and $y$ are prefix comparable. i.e. given a state symbol pair the previous and the next state both can be uniquely determined.

A transition in a reversible finite automaton can be defined as follows:

For $x_1, u_1, w_1$ where $x_1, w_1 \in V^*$ and $u_1 \in V \cup \{\lambda\}$ such that $x_1 u_1 w_1$ is the word on the input tape of the automaton and $q,q' \in Q$, $x_1 q u_1 w_1 \Rightarrow x_1 u_1 q' w_1$ iff there is transition rule $\delta(q,u_1)=q'$ in $\delta$. The symbol $\stackrel{*}{\Rightarrow}$ denotes the transitive and reflexive closure of $\Rightarrow$. The acceptance condition of reversible finite automaton is that the automaton needs to halt in a final state; it need not parse the complete input string. An automaton halts in a state if there is no transition defined for that current state and symbol pair or if the input string has been completely parsed.

The language accepted by a reversible finite automaton $M$ is $L(M)=\{w \in V^* | q_0 w \stackrel{*}{\Rightarrow} x_1 q u_1, \text{ with } q \in F, x_1 u_1 = w, x_1, u_1 \in V^*, \text{ and the automaton } M \text{ halts in state } q\}$.

### 2.2 Parallel communicating one-way reversible finite automata system

A parallel communicating one-way reversible finite automata system of degree $n$, denoted by PCRFA($n$), is a $(n+2)$-tuple $A=(V,A_1,A_2,...,A_n,K)$, where $V$ is the input alphabet, $A_i=(V,Q_i,q_i,F_i,\delta_i)$, $1 \leq i \leq n$, are reversible finite automata, where the sets $Q_i$ are not necessarily disjoint, $K=\{K_1,K_2,...,K_n\} \subseteq \cup_{i=1}^{n} Q_i$ is the set of query states. The automata $A_1,A_2,...,A_n$ are called the components of the system $A$. Note that any one-way reversible finite automaton is a parallel communicating one-way reversible finite automata system of degree $1$.

A configuration of a parallel communicating one-way reversible finite automata system is a $2n$-tuple $(s_1,u_1,s_2,u_2,...,s_n,u_n)$ where $s_i$ is the current state of the component $i$ and $u_i$ is the part of the input word which has not been read yet by the component $i$, for all $1 \leq i \leq n$. We define a binary relation $\Rightarrow$ on the set of all configurations by setting

$(s_1,u_1,s_2,u_2,...,s_n,u_n) \Rightarrow (r_1,u_1',r_2,u_2',...,r_n,u_n')$

if and only if one of the following two conditions holds:
1. $K \cap \{s_1 s_2,..., s_n\} = \emptyset$, $u_i = x_i u_i'$, $x_i \in V$ and $\delta(s_i,x_i)=r_i$, $1 \leq i \leq n$;
2. for all $1 \leq i \leq n$ such that $s_i = K_{j_i}$ where $K_{j_i}$ is a query state in $Q_i$ and $s_{j_i} \notin K$ is a state in $Q_i$ and also in $Q_j$ we have $r_i = s_{j_i}$, whereas for all the other $1 \leq z \leq n$ we have $r_z = s_z$. In this case $u_i' = u_i$, for all $1 \leq i \leq n$.

We denote by $\stackrel{*}{\Rightarrow}$ the reflexive and transitive closure of $\Rightarrow$, then the language recognized by a parallel communicating one-way reversible finite automata system is defined as:

$L(A)=\{w \in V^* | (q_1,w,q_2,w,...,q_n,w) \stackrel{*}{\Rightarrow} (s_1,u_1,s_2,u_2,...,s_n,u_n), s_i \in F_i, u_1,u_2,...,u_n \text{ are suffix of } w, \text{ and the reversible automaton } A_i \text{ halts in state } s_i, 1 \leq i \leq n\}$.

Intuitively, the language accepted by such a system consists of all words $w$ such that every component halts in a final state.

### 3. COMPUTATIONAL COMPLEXITY OF PARALLEL COMMUNICATING ONE-WAY REVERSIBLE FINITE AUTOMATA SYSTEM

In this Section, we discuss the computational power of parallel communicating one-way reversible finite automata system. We show that for every fully defined deterministic finite automaton which accepts a language $L$, we can construct a parallel communicating one-way reversible finite automata system which accepts the same language $L$. In order to explain the construction, we first employ the rules of construction on a particular deterministic finite automaton which accepts the language $(a+b)^*a$ (from Pin's work [5], we know that $L$ cannot be recognised by a one way reversible finite automaton with one head) and obtain the corresponding parallel communicating one-way reversible finite automata system which accepts the same language (See Example 1). Then in Theorem 1, we state the general proof for any deterministic finite automaton.

**Example 1:** Consider a fully defined deterministic finite automaton $M=(Q,V,q_0,F,\delta)$ which accepts the language $(a+b)^*a$, where $Q=\{q_0,q_1\}$, $V=\{a,b\}$, $F=\{q_1\}$. The transition function $\delta$ is as follows:

$\delta(q_0,a)=q_1$, $\delta(q_0,b)=q_0$, $\delta(q_1,a)=q_1$, $\delta(q_1,b)=q_0$

From Figure 1 it is clear that $M$ is not reversible. The corresponding centralized parallel communicating one-way reversible finite automata system $A$ that accepts $L=(a+b)^*a$ is obtained as follows:

Identify the states in the graphical representation of $M$ which are not reversible, i.e. in these states the number of in-degrees for a particular symbol is more than 1. For each such symbol and state pair $(s,a)$ identify the transitions of the form $\delta(q,a)=s$ in $M$ where $q \in Q$ and arrange all these transitions in an arbitrary order.

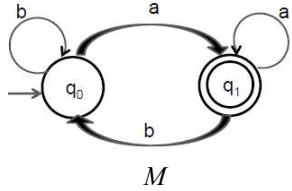

Figure 1: The DFA $M$ for the language $L=(a+b)^*a$, from the DFA it is evident that it's not reversible.

For our particular example both $q_0$ and $q_1$ are not reversible. The symbol and state pair responsible in making $q_0$ and $q_1$ not reversible are $(q_0,b)$ and $(q_1,a)$. The transitions associated with the pair $(q_0,b)$ are $\delta(q_0,b)=q_0$, $\delta(q_1,b)=q_0$ and the transitions associated with the pair $(q_1,a)$ are $\delta(q_0,a)=q_1$, $\delta(q_1,a)=q_1$.

Arrange and number these transitions in any arbitrary order from 2 to $n$, where $n$ is equal to the total no. of transitions in the list of transitions+1. For our particular example $n=5$ and the list of transitions are as follows:
2) $\delta(q_0,b)=q_0$
3) $\delta(q_1,b)=q_0$
4) $\delta(q_0,a)=q_1$
5) $\delta(q_1,a)=q_1$

For each transition $\delta(q_m,a)=q_j$ in the above list having number $k$ associated with it introduce a new component $A_k$ in the parallel communicating one-way reversible finite automata system. $A_k(\{q_i\},V,q_i,\{q_i\},\delta)$ and $\delta$ contains the transitions $\delta(q_i,x)=q_i$ for all $x \in V$. The new component $A_k$ is reversible.

Thus the new components for our particular example are as follows:
$A_2(\{q_0\},V,\{q_0\},\{\delta(q_0,a)=q_0, \delta(q_0,b)=q_0\})$,
$A_3(\{q_0\},V,\{q_0\},\{\delta(q_0,a)=q_0, \delta(q_0,b)=q_0\})$,
$A_4(\{q_1\},V,\{q_1\},\{\delta(q_1,a)=q_1, \delta(q_1,b)=q_1\})$,
$A_5(\{q_1\},V,\{q_1\},\{\delta(q_1,a)=q_1, \delta(q_1,b)=q_1\})$.

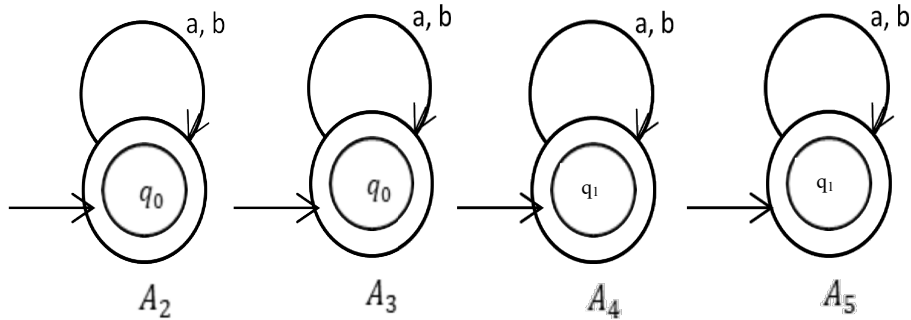

Figure 2: Components $A_2$, $A_3$, $A_4$ and $A_5$ of the parallel communicating one-way reversible finite automata system $A$.

It is evident from Figure 2 that $A_2$, $A_3$, $A_4$ and $A_5$ are all reversible.
Finally we design component 1, the construction of which depends on the deterministic finite automaton $M$.
$A_1=(Q,V,q_0,F,\delta')$.
For every transition in $\delta$
If a transition $\delta(q_m,a)=q_j$ is not in the list of transition created above then include $\delta(q_m,a)=q_j$ in $\delta'$.
If a transition $\delta(q_m,a)=q_j$ is in the list of transitions and it is in the $j^{th}$ position in the list then include $\delta(q_m,a)=K_j$ in $\delta$.
Therefore, $\delta'$ of component $A_1$ for our particular example is as follows:
$\delta(q_0,b)=K_2$, $\delta(q_1,b)=K_3$, $\delta(q_0,a)=K_4$, $\delta(q_1,a)=K_5$.

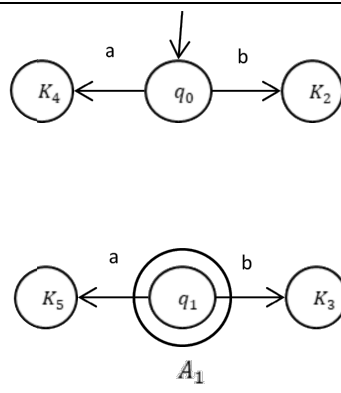

Figure 3: Component $A_1$ of the parallel communicating one-way reversible finite automata system $A$.

From Figure 3 it is evident that $A_1$ is also reversible. The set of query states $K=\{K_1,K_2,K_3,...,K_n\}$ one for each new component introduced in the construction. For our particular example $K=\{K_1,K_2,K_3,...,K_n\}$. Thus the parallel communicating one-way reversible finite automata system that is constructed from $M$ i.e. $A=(V,A_1,A_2,...,A_n,K)$, for our particular example $A=(V,A_1,A_2,A_3,A_4,A_5,\{K_1,K_2,K_3,K_4,K_5\})$.

**Theorem 1**: For every deterministic finite automaton $M$ which accepts a Language $L$ there exists a parallel communicating one-way reversible finite automata system $A=(V,A_1,A_2,...,A_n,K)$ which accepts the same language $L$.

**Proof:** Given a deterministic finite automaton $M=(Q,V,q_0,F,\delta)$ which accepts a language $L$, we can construct a parallel communicating one-way reversible finite automata system $A=(V,A_1,A_2,...,A_n,K)$ using the method described in Example 1.

Now consider a string $w \in L$, $M$ either accepts $w$ or rejects $w$, but in doing so $M$ goes through a sequence of transitions and states. As $M$ is a deterministic finite automaton therefore $M$ accepts only after completely consuming the input, thus the length of both the sequence of transitions and states is $|w|$. Parallel communicating one-way reversible finite automata system $A$ imitates the sequence of transitions of $M$ using the component $A_1$.

1) If a transition $\delta(q_m,x)=q_i$ in the sequence of transitions is not in the list of transitions constructed above then the according to the construction, the transition function $\delta_i$ contains the transition and it implements it. Thus after implementation of the transition both $M$ and $A_i$ are in the same state.
2) If a transition $\delta(q_m,x)=q_i$ in the sequence of transitions is in the list of transitions constructed above then that transition is associated with a number m. $A_1$ in order to simulate $\delta(q_m,x)=q_i$ uses the transition $\delta(q_m,x)=K_m$ which is the query state of component $m$. According to the construction, component $m$ only contains state $q_i$ which is reported back to $A_1$ and $A_1$ goes to state $q_i$. Thus after implementation of the transition both $M$ and $A_i$ are in the same state.

The sequence of states through which $A_1$ and $M$ goes on input $w$ are the same. If $M$ accepts $w$ by going to its final state after consuming $w$ so does $A_1$. Other components of $A$ contain single state which is a final state so they also accept $w$ therefore $A$ accepts $w$. If $M$ rejects $w$ so does $A_1$, and although all other components of $A$ accept $w$ but as $A_1$ does not accepts $w$, $A$ rejects $w$.

Thus both $M$ and $A$ accept the same language $L$.

Corollary 1 follows from Theorem 1.

**Corollary 1:** Centralized parallel communicating one-way reversible finite automata systems can accept all regular languages.

**Theorem 2:** For every one-way reversible multi-head finite automaton $M$ there exists a parallel communicating one-way reversible finite automata system $A$ that accepts the same language as $M$.

**Proof:** The proof of this theorem is same as the proof in mitrana et.al.[6] which shows for every one-way deterministic multi-head finite automaton $M$ there exists a deterministic parallel communicating deterministic finite automata system $A$ that accepts the same language as $M$. Let $M=(Q,V,k,\#,\$,q_0,F,\delta)$ be a multi-head reversible finite automaton. A multi-head reversible finite automaton is a reversible finite automaton with multiple-heads denoted by $k$; the transitions are so defined that one can uniquely determine the previous state and the next state given the current state and the k symbols read by the k input heads. We construct the parallel communicating one-way reversible finite automata system $A=(V,A_1,A_2,...,A_n,K)$ in the following manner:
$A_i=(Q_i,V,q_0,F,\delta_i)$, where
$Q_i=K\cup Q\cup(Qx(V\cup\{\lambda\})^{i-1}x(V\cup\{\lambda\})^{i-1})\cup(Qx(V\cup\{\lambda\})^i x(V\cup\{\lambda\})^i)\cup X_i\cup Y_i$
with

$X_i = \begin{cases} \emptyset, & i \leq 2 \\ \{p_i|, 1 \leq i \leq i-2\}, & i > 2, \end{cases}$ for e.g. if $i=5$, then $X_5=\{p_1, p_2, p_3\}$ these states in $X_i$ allow the automaton $A_i$ in $A$ to wait for the previous automata i.e. $A_j, 1\leq j<i$ to complete their transitions.

$$Y_i = \begin{cases} \emptyset, & i = n \\ \{s_j | i+1 \leq j \leq n\}, & i < n, \end{cases}$$ for e.g. if $i=5$, $n=8$ then $Y_5=\{s_6, s_7, s_8\}$ these states in $Y_i$ allow the automaton $A_i$ in $A$ to wait for the later automata i.e. $A_k, 1 \leq k < i$ to complete their transitions.

So that all the automata in $A$ can go to a state $q \in Q$ at the same time.

For a transition $\delta(q_i, a_1, a_2, \ldots, a_m) = q_j$, $a_k \in V \cup \{\lambda\}$, $1 \leq k \leq m$, and $q_i, q_j \in Q$.

The transitions introduced in the component $A_i$ of $A$ i.e. in $\delta_i$, $1 \leq i \leq n$ are as follows:

All the components begin in state $q_i$, all the components are in waiting except the first component. The first component reads input from the input tape based on $a_1$ read by the multi-head reversible finite automaton and stores the symbol in its current state. The control then switches to component two, and the current state of component one is passed to component two. All the other components are still in waiting.

$i=1$: $\delta_1(q_i, a_1) = (q_i, a_1)$,
$\delta_1((q_i, a_1), \lambda) = s_2$,
$\delta_1(s_j, \lambda) = s_{j+1}$, $2 \leq j \leq n-1$,
$\delta_1(s_n, \lambda) = K_n$

Components 2 to n-1 behave in the same manner as component one that is they read the symbols according to transition of the multi-head reversible finite automaton the i$^{th}$ components head reads the symbol $a_i$ and stores the symbol read along with the symbol read information it received from the component before it in its current state. It switches the control to its next component and also passes its current state to the next state.

$i=2$: $\delta_2(q_i, \lambda) = K_1$
$\delta_2((q_i, a_1), a_2) = (q_i, a_1, a_2)$,
$\delta_2((q_i, a_1, a_2), \lambda) = s_3$,
$\delta_2(s_j, \lambda) = s_{j+1}$, $3 \leq j \leq n-1$,
$\delta_2(s_n, \lambda) = K_n$

$i=2 < i < n$ $\delta_i(q_i, \lambda) = p_1$
$\delta_i(p_j, \lambda) = p_{j+1}$, $1 \leq j \leq i-3$
$\delta_i(p_{i-2}, \binom{\lambda}{\lambda}) = K_{i-1}$
$\delta_i((q_i, a_1, a_2, \ldots, a_{i-1}), a_i) = (q_i, a_1, a_2, \ldots, a_i)$,
$\delta_i((q_i, a_1, a_2, \ldots, a_i), \lambda) = s_{i+1}$,
$\delta_i(s_j, \lambda) = s_{j+1}$, $i+1 \leq j \leq n-1$,
$\delta_i(s_n, \lambda) = K_n$

In the n$^{th}$ component, its head reads the symbol $a_n$ and $b_n$, all other components are waiting and then it goes to state $q_j$, and simultaneously sends its current state information to all other components. Thus, all other components also go to state $q_j$.

$i=n$ $\delta_n(q_i, \lambda) = p_1$
$\delta_n(p_j, \lambda) = p_{j+1}$, $1 \leq j \leq n-3$
$\delta_n(p_{n-2}, \lambda) = K_{n-1}$
$\delta_n((q_i, a_1, a_2, \ldots, a_{n-1}), a_n) = (q_i, a_1, a_2, \ldots, a_n)$,
$\delta_n((q_i, a_1, a_2, \ldots, a_n), \lambda) = q_j$,

Thus, the parallel communicating one-way reversible finite automata system simulates one transition of the multi-head reversible finite automaton in the above stated manner. Thus any transition the multi-head reversible finite automaton makes, the parallel communicating one-way reversible finite automata system can replicate it. Moreover if at a particular instance the multi-head reversible finite automaton does not have a transition defined and rejects the input, a similar behaviour is expected from the parallel communicating one-way reversible finite automata system as at least one of its component will also not have transition defined as a result that component will halt before others and thus reject the input also.

**Observation 1:** Even though the components of the parallel communicating one-way reversible finite automata system are reversible the system as a whole is not reversible given a configuration of the system and the symbol read by the input head of each component the previous configuration cannot be uniquely determined. For e.g. in Example 1 if the system $A$ has state configuration $(q_1, q_0, q_0, q_1, q_1)$ and input $(a,a,a,a,a)$ read by the heads of the components the previous state can be $(q_1, q_0, q_0, q_1, q_1)$ or $(q_0, q_0, q_0, q_1, q_1)$ as both query states $K_4$ and $K_5$ goes to state $q_1$ so the state $q_1$ of component 1 can be reached from either $q_0$ or $q_1$ on input $a$.

**Conjecture 1:** Parallel communicating one-way reversible finite automata systems are computationally more powerful than one-way reversible multi-head finite automaton.

The conjecture is based on Theorem 2 and Observation 1. The system can be non-reversible in spite of the components being reversible. Therefore, there is a high chance that parallel communicating one-way reversible finite automata systems may accept some non reversible languages.

## 4. CONCLUSION

In this paper, we introduce a model of one-way reversible finite automaton and show that the new model namely parallel communicating one-way reversible finite automata system can accept all regular languages. We further show that this new model accepts all languages which are accepted by multi-head reversible finite automata. Moreover, we make an interesting observation that even though the components of reversible finite automata system are reversible the system as a whole is not reversible, which is unlike other parallel communicating models. From the above observation we make an interesting conjecture that parallel communicating one-way reversible finite automata systems may be more powerful than multi-head reversible finite automata.